\documentstyle[12pt]{article}
\def\ga{\mathrel{\mathpalette\fun >}}
\def\fun#1#2{\lower3.6pt\vbox{\baselineskip0pt\lineskip.9pt
\ialign{$\mathsurround=0pt#1\hfil##\hfil$\crcr#2\crcr\sim\crcr}}}
\newcommand{\dd}{\mbox{d}}

\begin{document}

\begin{titlepage}
\rightline{CERN-TH/95-241}
\rightline{JINR-E2-95-110}

\noindent
\begin{center}
{\huge \bf Pair Production in Small-Angle \\ Bhabha Scattering }
\end{center}
%\vspace{1cm}
\bigskip

\begin{center}
{\large A.B.~Arbuzov,  E.A.~Kuraev\\
\vspace{0.2cm}
Bogoliubov Laboratory of Theoretical Physics, JINR,\\ Dubna, 141980,
Russia\\
\vspace{0.5cm}
N.P.~Merenkov\\
\vspace{0.2cm}
Physico--Technical Institute,
Kharkov, 310108, Ukraine
\\
\vspace{0.2cm}
and\\
\vspace{0.2cm}
L.~Trentadue \footnote{\footnotesize On leave of absence from the
Dipartimento di Fisica, Universit\'a di Parma, Parma, Italy.}$^,$
\footnote{\footnotesize INFN, Gruppo Collegato di Parma, Sezione di Milano,
Milano, Italy.}\\
\vspace{0.2cm}
\noindent
Theoretical Physics Division, CERN,\\ CH-1211 Geneva 23, Switzerland\\}
\end{center}

\medskip\noindent
\bigskip
\begin{center}
{\bf Abstract} \\
\end{center}
{ Radiative corrections due to the production
of virtual as well as
real soft and hard pairs in small-angle
Bhabha scattering are calculated analytically.
Both the collinear and the semi-collinear kinematical regions
of hard pair production are considered.
The calculation, within the leading and next-to-leading logarithmic
approximation, provides an accuracy of $O(0.1\%)$.
Numerical results show that the effects of pair production have
to be taken into account for a luminosity determination accurate to
$O(0.1\%)$ at LEP.}

\vskip 12pt \noindent

\vfil
\leftline{CERN-TH/95-241}
\leftline{September 1995}
\end{titlepage}

\noindent

\section{Introduction}

The electron--positron scattering process (Bhabha process) at small
angles is used for the measurement of the luminosity
at LEP~I [1]. This technique provides
an experimental accuracy of the $O(0.1\%)$, or even better [2].
Such an accurate theoretical calculation of the Bhabha cross-section
was missing [3]. Recently [4] the radiative corrections due to the
emission of virtual, real soft and hard photons and pairs have been
calculated up to the three loop providing an $O(0.1\%)$ accurate
cross-section. In this work [4] we gave the
results of the
performed analytical calculations. The leading $(\sim (\alpha L/\pi)^{1,2,3})$
contributions as well as the next to leading
$(\sim \alpha/\pi,\ (\alpha /\pi)^{2}L)$ ones were calculated explicitly
for the processes with the emission of photons where
$L=\ln Q^2/m_e^2\ $ and $Q^2\sim 10$ $(\mbox{GeV}/c)^2$ is
the squared momentum transferred. For a given scattering angle $\theta$
one has $Q^2=2\epsilon^2(1-cos\;\theta)$ with $\epsilon$ the beam energy.
Also the pair production processes and the contributions due to the
emission of virtual, soft and real hard pairs were considered. However,
the production of real hard pairs was calculated only in the collinear
kinematics (CK) limit [4]. In order to assess the accuracy already obtained
within the collinear limit, in this paper, we carry a systematic study of the
hard pair emission within the semi-collinear kinematics (SCK). We present also
the total
contribution to the observable Bhabha cross-section due to the pair
production
\begin{eqnarray} \label{eq:1}
e^-(p_1)\ +\ e^+(p_2) \rightarrow e^-(q_1)\ +\ e^+(q_2)\
+\ e^-(p_-)\ +\ e^+(p_+)\, ,
\end{eqnarray}
which takes into account the cuts on the detection of the scattered
electron and positron. We accept the convention [1--3] to consider
as an event of the Bhabha process one in which the angles of the
simultaneously registered particles hitting the opposite detectors
lay in the range:
\begin{eqnarray} \label{eq:2}
\theta_{\small \mbox{min}}<\theta_e<\theta_{\mbox{\small max}}
=\rho\theta_{\mbox{\small min}},
\qquad \pi-\rho\theta_{\mbox{\small min}}<\theta_{\bar{e}}<\pi
-\theta_{\mbox{\small min}},
\end{eqnarray}
$(\theta_{\mbox{\small min}}\sim 3^{\circ}\ $, $\rho \ga 1)$ with respect to
the beam direction. The second condition is imposed on the energy fractions
of the scattered electron and positron:
\begin{eqnarray} \label{eq:3}
x_ex_{\bar{e}} > x_c,\qquad x_{e,\bar{e}}
=2\varepsilon_{e,\bar{e}}/\sqrt{s}, \qquad s=4\varepsilon^2,
\end{eqnarray}
where $\varepsilon$ is the energy of the initial electron (or positron),
$\varepsilon_{e,(\bar{e})}$ is the energy of the scattered electron
(positron)\footnote{\footnotesize Here and in the
following, it is implied that the centre of mass is the reference frame.}.

Our method for the real hard pair production cross-section calculation
within the logarithmic accuracy consists in the separation of the
contributions due to the collinear and semi-collinear kinematical regions
[5,6].
In the first one (CK)  we suggest that both electron and positron from
the created pair go in the narrow cone along to the direction of one
of the charged particles (the projectile (scattered) electron $\vec{p}_1$
$(\vec{q}_1)$ or the projectile (scattered) positron $\vec{p}_2$
$(\vec{q}_2))$:
\begin{eqnarray} \label{eq:4}
\widehat{\vec{p}_+\vec{p}_-} \sim \widehat{\vec{p}_-\vec{p}_i}
\sim \widehat{\vec{p}_+\vec{p}_i}
< \theta_0 \ll 1, \qquad \varepsilon\theta_0/m \gg 1, \qquad
\vec{p}_i=\vec{p}_1,\,\vec{p}_2,\,\vec{q}_1,\,\vec{q}_2\, .
\end{eqnarray}
The contribution of the CK contains terms of the order of
$(\alpha L/\pi)^2$ and $(\alpha \pi)^2L$. In the semi-collinear region
only one of conditions (\ref{eq:4}) on the angles is fulfilled:
\begin{eqnarray} \label{eq:5}
&& \widehat{\vec{p}_+\vec{p}_-} < \theta_0,\quad
\widehat{\vec{p}_{\pm}\vec{p}_i} > \theta_0\, ; \quad \mbox{or} \quad
\widehat{\vec{p}_-\vec{p}_i} < \theta_0,\quad
\widehat{\vec{p}_+\vec{p}_i} > \theta_0\, ; \\ \nonumber
&& \mbox{or} \quad
\widehat{\vec{p}_-\vec{p}_i} > \theta_0,\quad
\widehat{\vec{p}_+\vec{p}_i} < \theta_0\, .
\end{eqnarray}
The contribution of the SCK contains the terms of the form:
\begin{eqnarray} \label{eq:6}
\left( \frac{\alpha}{\pi} \right) ^2 L \ln\frac{\theta_0}{\theta},
\qquad \left( \frac{\alpha}{\pi} \right) ^2 L,
\end{eqnarray}
where $\theta = \widehat{\vec{p}_-\vec{q}_1}$ is the scattering angle.
The auxiliary parameter $\theta_0$ disappears in the total sum of
the CK and SCK contributions. We systematically omit the
terms without large logarithms; they are of the form
$(\alpha/\pi)^2\cdot\mbox{const}\sim 10^{-5}$.

We restrict ourselves to the case when the electron--positron pair is
created. The effects due to the other pair creation $(\mu^+\mu^-$,
$\pi^+\pi^-$, etc.) are at least one order of magnitude smaller and can be
neglected, as will be seen from the numerical analysis.
All possible mechanisms for pair creation (the singlet and
non-singlet ones) as well as the identity of the particles in
the final state are taken into account. In the case of small-angle
Bhabha scattering, only the scattering-type diagrams are relevant
among the the total 36 Feynman tree level diagrams. Besides,
we are convinced of the cancellations of the interference between the
amplitudes describing the production of pairs moving along the electron
direction and the positron one known as up--down
cancellation.

The sum of the contributions due to the virtual pair emission
(due to the vacuum polarization insertions in the virtual photon
Green function) and of those due to the real soft pair emission
does not contain cubic $(\sim L^3)$ terms, but depends on the auxiliary
parameter $\Delta = \delta\varepsilon/\varepsilon$
$(m_e\ll\delta\varepsilon\ll\varepsilon$, $\ \varepsilon$ is the energy sum
of the soft pair components).
The $\Delta$-dependence disappears in the total sum after adding
the contributions due to the real hard pair production. Before summing
one has to integrate the hard pair contributions over the energy
fractions of pair components as well as over the ones of the scattered
electron and positron:
\begin{eqnarray} \label{eq:7}
\Delta&=&\frac{\delta\varepsilon}{\varepsilon}< x_1+x_2, \qquad
x_c<x=1-x_1-x_2<1-\Delta, \\ \nonumber
x_1&=&\frac{\varepsilon_+}{\varepsilon}, \qquad
x_2=\frac{\varepsilon_-}{\varepsilon}, \qquad
x  =\frac{q_1^0}{\varepsilon}\, ,
\end{eqnarray}
where $\varepsilon_{\pm}$ are the energies of the positron and electron from
the created pair. We consider for definiteness the case where the created hard
pair moves close to the direction of the initial (or scattered) electron.

The paper is organized as follows: in the second part we consider
the emission of the hard pair in the collinear kinematics. The results
turned out to be very close to the ones obtained by one of us (N.P.M.)
in paper [6] for the case of the pair production in electron--nuclei
scattering and applied to the case of small-angle Bhabha scattering in [4].
For completeness, we present very briefly the derivation and
give the result correcting some misprints in [6]. In the third part
we consider the semi-collinear kinematical regions. The differential
cross-section is obtained there and integrated over the angles and
energy fractions of the pair components.
In the fourth part we give the expression of the radiative corrections
to the experimental cross-section due to pair production.
The results are illustrated numerically in tables and discussed in the
Conclusions.

\section{The Collinear Kinematics}

In evaluating the cross-section we see that there are four different
CK regions: when the created pair goes
along the direction of the initial (scattered) electron or positron [4].
We will consider only two of them, corresponding to the initial and
the final electron directions. For the case of the pair emission
along the initial electron, it is useful to decompose the particle
momenta into the longitudinal and transverse components:
\begin{eqnarray} \label{eq:8}
p_+&=&x_1p_1 + p_+^{\bot}, \qquad p_-=x_2p_1 + p_-^{\bot},
\qquad q_1=xp_1 + q_1^{\bot}, \\ \nonumber
x&=&1-x_1-x_2, \qquad q_2 \approx p_2, \qquad
p_+^{\bot} + p_-^{\bot} + q_1^{\bot}=0,
\end{eqnarray}
where $p_i^{\bot}$ are the transverse
two-dimensional momenta of the final
particles with respect to the initial
electron beam direction . It is convenient to introduce the following
dimensionless
quantities for the relevant kinematical invariants:
\begin{eqnarray} \label{eq:9}
z_i&=&\left(\frac{\varepsilon\theta_i}{m}\right)^2, \quad
z_1=\left(\frac{p_-^{\bot}}{m}\right)^2, \quad
z_2=\left(\frac{p_+^{\bot}}{m}\right)^2, \quad
0<z_i< \left(\frac{\varepsilon\theta_0}{m}\right)^2 \gg 1, \\ \nonumber
A&=&\frac{(p_++p_-)^2}{m^2}=(x_1x_2)^{-1}\bigl[
(1-x)^2+x_1^2x_2^2(z_1+z_2-2\sqrt{z_1z_2}\cos\phi )\bigr], \\ \nonumber
A_1&=&\frac{2p_1p_-}{m^2}=x_2^{-1}\bigl[1+x_2^2+x_2^2z_2\bigr], \qquad
A_2=\frac{2p_1p_+}{m^2}=x_1^{-1}\bigl[1+x_1^2+x_1^2z_1\bigr], \\ \nonumber
C&=&\frac{(p_1-p_-)^2}{m^2}=2-A_1,
\qquad D=\frac{(p_1-q_1)^2}{m^2}-1=A-A_1-A_2,
\end{eqnarray}
where $\phi$ is the azimuthal angle between the planes
$(\vec{p}_1 p_+^{\bot})$ and $(\vec{p}_1 p_-^{\bot})$.

Keeping only the terms from the squared matrix element module summed over the
spin states module that give non zero contribution
to the cross-section in the limit $\theta_0 \to 0$, we
find that only 8 of the 36 tree level Feynman diagrams
are essential. They are drawn in fig.~1.

%-----------------fig.1-----------------------------------
%\begin{figure}
\vspace{.3cm}
\unitlength=.7mm
\special{em:linewidth 0.4pt}
\linethickness{0.4pt}
\begin{picture}(199.00,90.00)
\put(10.00,72.00){\vector(1,0){7.00}}
\put(17.00,72.00){\line(1,0){6.00}}
\put(27.00,72.00){\vector(1,0){12.00}}
\put(39.00,72.00){\line(1,0){6.00}}
\put(45.00,60.00){\vector(-1,0){7.00}}
\put(38.00,60.00){\vector(-1,0){23.00}}
\put(10.00,60.00){\line(1,0){5.00}}
\put(30.00,61.50){\oval(3.00,3.00)[l]}
\put(30.00,64.50){\oval(3.00,3.00)[r]}
\put(30.00,67.50){\oval(3.00,3.00)[l]}
\put(30.00,70.50){\oval(3.00,3.00)[r]}
\put(23.00,72.00){\oval(4.00,4.00)[lt]}
\put(23.00,76.00){\oval(4.00,4.00)[rb]}
\put(27.00,76.00){\oval(4.00,4.00)[lt]}
\put(27.00,78.00){\vector(2,1){7.00}}
\put(34.00,81.50){\line(2,1){6.00}}
\put(40.00,78.00){\vector(-1,0){7.00}}
\put(27.00,78.00){\line(1,0){6.00}}
\put(20.00,72.00){\line(1,0){10.00}}
\put(9.00,76.00){\makebox(0,0)[cc]{$p_1$}}
\put(9.00,56.00){\makebox(0,0)[cc]{$p_2$}}
\put(34.00,65.00){\makebox(0,0)[cc]{$q$}}
\put(44.00,86.00){\makebox(0,0)[cc]{$p_-$}}
\put(44.00,79.00){\makebox(0,0)[cc]{$p_+$}}
\put(49.00,73.00){\makebox(0,0)[cc]{$q_1$}}
\put(47.00,56.00){\makebox(0,0)[cc]{$q_2$}}
\put(53.00,66.00){\makebox(0,0)[cc]{$+$}}

\put(60.00,72.00){\vector(1,0){7.00}}
\put(67.00,72.00){\line(1,0){6.00}}
\put(77.00,72.00){\vector(1,0){12.00}}
\put(89.00,72.00){\line(1,0){6.00}}
\put(95.00,60.00){\vector(-1,0){7.00}}
\put(88.00,60.00){\vector(-1,0){23.00}}
\put(60.00,60.00){\line(1,0){5.00}}
\put(72.00,61.50){\oval(3.00,3.00)[l]}
\put(72.00,64.50){\oval(3.00,3.00)[r]}
\put(72.00,67.50){\oval(3.00,3.00)[l]}
\put(72.00,70.50){\oval(3.00,3.00)[r]}
\put(78.00,72.00){\oval(4.00,4.00)[lt]}
\put(78.00,76.00){\oval(4.00,4.00)[rb]}
\put(82.00,76.00){\oval(4.00,4.00)[lt]}
\put(82.00,78.00){\vector(2,1){7.00}}
\put(89.00,81.50){\line(2,1){6.00}}
\put(95.00,78.00){\vector(-1,0){7.00}}
\put(82.00,78.00){\line(1,0){6.00}}
\put(70.00,72.00){\line(1,0){10.00}}
\put(62.00,77.00){\makebox(0,0)[cc]{$p_1$}}
\put(62.00,56.00){\makebox(0,0)[cc]{$p_2$}}
\put(99.00,73.00){\makebox(0,0)[cc]{$q_1$}}
\put(99.00,56.00){\makebox(0,0)[cc]{$q_2$}}
\put(99.00,80.00){\makebox(0,0)[cc]{$p_+$}}
\put(99.00,86.00){\makebox(0,0)[cc]{$p_-$}}
\put(103.00,66.00){\makebox(0,0)[cc]{$-$}}

\put(110.00,72.00){\vector(1,0){7.00}}
\put(117.00,72.00){\line(1,0){6.00}}
\put(127.00,72.00){\vector(1,0){12.00}}
\put(139.00,72.00){\line(1,0){6.00}}
\put(145.00,60.00){\vector(-1,0){7.00}}
\put(138.00,60.00){\vector(-1,0){23.00}}
\put(110.00,60.00){\line(1,0){5.00}}
\put(130.00,61.50){\oval(3.00,3.00)[l]}
\put(130.00,64.50){\oval(3.00,3.00)[r]}
\put(130.00,67.50){\oval(3.00,3.00)[l]}
\put(130.00,70.50){\oval(3.00,3.00)[r]}
\put(123.00,72.00){\oval(4.00,4.00)[lt]}
\put(123.00,76.00){\oval(4.00,4.00)[rb]}
\put(127.00,76.00){\oval(4.00,4.00)[lt]}
\put(127.00,78.00){\vector(2,1){7.00}}
\put(134.00,81.50){\line(2,1){6.00}}
\put(140.00,78.00){\vector(-1,0){7.00}}
\put(127.00,78.00){\line(1,0){6.00}}
\put(120.00,72.00){\line(1,0){10.00}}
\put(109.00,76.00){\makebox(0,0)[cc]{$p_1$}}
\put(109.00,56.00){\makebox(0,0)[cc]{$p_2$}}
\put(144.00,86.00){\makebox(0,0)[cc]{$q_1$}}
\put(144.00,79.00){\makebox(0,0)[cc]{$p_+$}}
\put(149.00,73.00){\makebox(0,0)[cc]{$p_-$}}
\put(147.00,56.00){\makebox(0,0)[cc]{$q_2$}}
\put(153.00,66.00){\makebox(0,0)[cc]{$-$}}

\put(160.00,72.00){\vector(1,0){7.00}}
\put(167.00,72.00){\line(1,0){6.00}}
\put(177.00,72.00){\vector(1,0){12.00}}
\put(189.00,72.00){\line(1,0){6.00}}
\put(195.00,60.00){\vector(-1,0){7.00}}
\put(188.00,60.00){\vector(-1,0){23.00}}
\put(160.00,60.00){\line(1,0){5.00}}
\put(172.00,61.50){\oval(3.00,3.00)[l]}
\put(172.00,64.50){\oval(3.00,3.00)[r]}
\put(172.00,67.50){\oval(3.00,3.00)[l]}
\put(172.00,70.50){\oval(3.00,3.00)[r]}
\put(178.00,72.00){\oval(4.00,4.00)[lt]}
\put(178.00,76.00){\oval(4.00,4.00)[rb]}
\put(182.00,76.00){\oval(4.00,4.00)[lt]}
\put(182.00,78.00){\vector(2,1){7.00}}
\put(189.00,81.50){\line(2,1){6.00}}
\put(195.00,78.00){\vector(-1,0){7.00}}
\put(182.00,78.00){\line(1,0){6.00}}
\put(170.00,72.00){\line(1,0){10.00}}
\put(162.00,77.00){\makebox(0,0)[cc]{$p_-$}}
\put(162.00,56.00){\makebox(0,0)[cc]{$p_2$}}
\put(199.00,73.00){\makebox(0,0)[cc]{$p_-$}}
\put(199.00,56.00){\makebox(0,0)[cc]{$q_2$}}
\put(199.00,79.00){\makebox(0,0)[cc]{$p_+$}}
\put(199.00,86.00){\makebox(0,0)[cc]{$q_1$}}

\put(45.00,10.00){\vector(-1,0){7.00}}
\put(38.00,10.00){\vector(-1,0){23.00}}
\put(10.00,10.00){\line(1,0){5.00}}
\put(12.00,6.00){\makebox(0,0)[cc]{$p_2$}}
\put(47.00,6.00){\makebox(0,0)[cc]{$q_2$}}
\put(22.00,11.50){\oval(3.00,3.00)[r]}
\put(22.00,14.50){\oval(3.00,3.00)[l]}
\put(10.00,30.00){\vector(1,0){7.00}}
\put(17.00,30.00){\line(1,0){6.00}}
\put(27.00,30.00){\vector(1,0){12.00}}
\put(39.00,30.00){\line(1,0){6.00}}
\put(22.00,25.50){\oval(3.00,3.00)[l]}
\put(22.00,28.50){\oval(3.00,3.00)[r]}
\put(20.00,30.00){\line(1,0){10.00}}
\put(12.00,35.00){\makebox(0,0)[cc]{$p_1$}}
\put(47.00,35.00){\makebox(0,0)[cc]{$q_1$}}
\put(22.00,24.00){\vector(1,0){8.00}}
\put(30.00,24.00){\line(1,0){5.00}}
\put(35.00,16.00){\vector(-1,0){7.00}}
\put(22.00,16.00){\line(1,0){6.00}}
\put(22.00,16.00){\line(0,1){8.00}}
\put(39.00,24.00){\makebox(0,0)[cc]{$p_-$}}
\put(39.00,16.00){\makebox(0,0)[cc]{$p_+$}}
\put(53.00,20.00){\makebox(0,0)[cc]{$+$}}

\put(94.00,10.00){\vector(-1,0){7.00}}
\put(87.00,10.00){\vector(-1,0){23.00}}
\put(59.00,10.00){\line(1,0){5.00}}
\put(61.00,6.00){\makebox(0,0)[cc]{$p_2$}}
\put(96.00,6.00){\makebox(0,0)[cc]{$q_2$}}
\put(71.00,11.50){\oval(3.00,3.00)[r]}
\put(71.00,14.50){\oval(3.00,3.00)[l]}
\put(59.00,30.00){\vector(1,0){7.00}}
\put(66.00,30.00){\line(1,0){6.00}}
\put(76.00,30.00){\vector(1,0){12.00}}
\put(88.00,30.00){\line(1,0){6.00}}
\put(71.00,25.50){\oval(3.00,3.00)[l]}
\put(71.00,28.50){\oval(3.00,3.00)[r]}
\put(69.00,30.00){\line(1,0){10.00}}
\put(61.00,35.00){\makebox(0,0)[cc]{$p_1$}}
\put(96.00,35.00){\makebox(0,0)[cc]{$q_1$}}
\put(71.00,16.00){\vector(1,0){8.00}}
\put(79.00,16.00){\line(1,0){5.00}}
\put(84.00,24.00){\vector(-1,0){7.00}}
\put(71.00,24.00){\line(1,0){6.00}}
\put(71.00,16.00){\line(0,1){8.00}}
\put(88.00,16.00){\makebox(0,0)[cc]{$p_-$}}
\put(88.00,24.00){\makebox(0,0)[cc]{$p_+$}}
\put(3.00,20.00){\makebox(0,0)[cc]{$+$}}
\put(103.00,20.00){\makebox(0,0)[cc]{$-$}}
\put(153.00,20.00){\makebox(0,0)[cc]{$-$}}

\put(27.50,50.00){\makebox(0,0)[cc]{(1)}}
\put(77.50,50.00){\makebox(0,0)[cc]{(2)}}
\put(127.50,50.00){\makebox(0,0)[cc]{(3)}}
\put(177.50,50.00){\makebox(0,0)[cc]{(4)}}
\put(27.50,0.00){\makebox(0,0)[cc]{(5)}}
\put(77.50,0.00){\makebox(0,0)[cc]{(6)}}
\put(127.50,0.00){\makebox(0,0)[cc]{(7)}}
\put(177.50,0.00){\makebox(0,0)[cc]{(8)}}

\put(145.00,10.00){\vector(-1,0){7.00}}
\put(138.00,10.00){\vector(-1,0){23.00}}
\put(110.00,10.00){\line(1,0){5.00}}
\put(112.00,6.00){\makebox(0,0)[cc]{$p_2$}}
\put(147.00,6.00){\makebox(0,0)[cc]{$q_2$}}
\put(122.00,11.50){\oval(3.00,3.00)[r]}
\put(122.00,14.50){\oval(3.00,3.00)[l]}
\put(110.00,30.00){\vector(1,0){7.00}}
\put(117.00,30.00){\line(1,0){6.00}}
\put(127.00,30.00){\vector(1,0){12.00}}
\put(139.00,30.00){\line(1,0){6.00}}
\put(122.00,25.50){\oval(3.00,3.00)[l]}
\put(122.00,28.50){\oval(3.00,3.00)[r]}
\put(120.00,30.00){\line(1,0){10.00}}
\put(112.00,35.00){\makebox(0,0)[cc]{$p_1$}}
\put(147.00,35.00){\makebox(0,0)[cc]{$p_-$}}
\put(122.00,24.00){\vector(1,0){8.00}}
\put(130.00,24.00){\line(1,0){5.00}}
\put(135.00,16.00){\vector(-1,0){7.00}}
\put(122.00,16.00){\line(1,0){6.00}}
\put(122.00,16.00){\line(0,1){8.00}}
\put(139.00,24.00){\makebox(0,0)[cc]{$q_1$}}
\put(139.00,16.00){\makebox(0,0)[cc]{$p_+$}}
\put(194.00,10.00){\vector(-1,0){7.00}}
\put(187.00,10.00){\vector(-1,0){23.00}}
\put(159.00,10.00){\line(1,0){5.00}}
\put(161.00,6.00){\makebox(0,0)[cc]{$p_2$}}
\put(196.00,6.00){\makebox(0,0)[cc]{$q_2$}}
\put(171.00,11.50){\oval(3.00,3.00)[r]}
\put(171.00,14.50){\oval(3.00,3.00)[l]}
\put(159.00,30.00){\vector(1,0){7.00}}
\put(166.00,30.00){\line(1,0){6.00}}
\put(176.00,30.00){\vector(1,0){12.00}}
\put(188.00,30.00){\line(1,0){6.00}}
\put(171.00,25.50){\oval(3.00,3.00)[l]}
\put(171.00,28.50){\oval(3.00,3.00)[r]}
\put(169.00,30.00){\line(1,0){10.00}}
\put(161.00,35.00){\makebox(0,0)[cc]{$p_1$}}
\put(196.00,35.00){\makebox(0,0)[cc]{$p_-$}}
\put(171.00,16.00){\vector(1,0){8.00}}
\put(179.00,16.00){\line(1,0){5.00}}
\put(184.00,24.00){\vector(-1,0){7.00}}
\put(171.00,24.00){\line(1,0){6.00}}
\put(171.00,16.00){\line(0,1){8.00}}
\put(188.00,16.00){\makebox(0,0)[cc]{$q_1$}}
\put(188.00,24.00){\makebox(0,0)[cc]{$p_+$}}
\end{picture}
%\caption{The Feynman diagrams giving logarithmically
%enhanced contributions in the kinematical region where the created
%pair goes along the electron direction. The signs represent the
%Fermi--Dirac statistics of the interchanged fermions.}
%\end{figure}

\nopagebreak
\vspace{.1cm}
\noindent
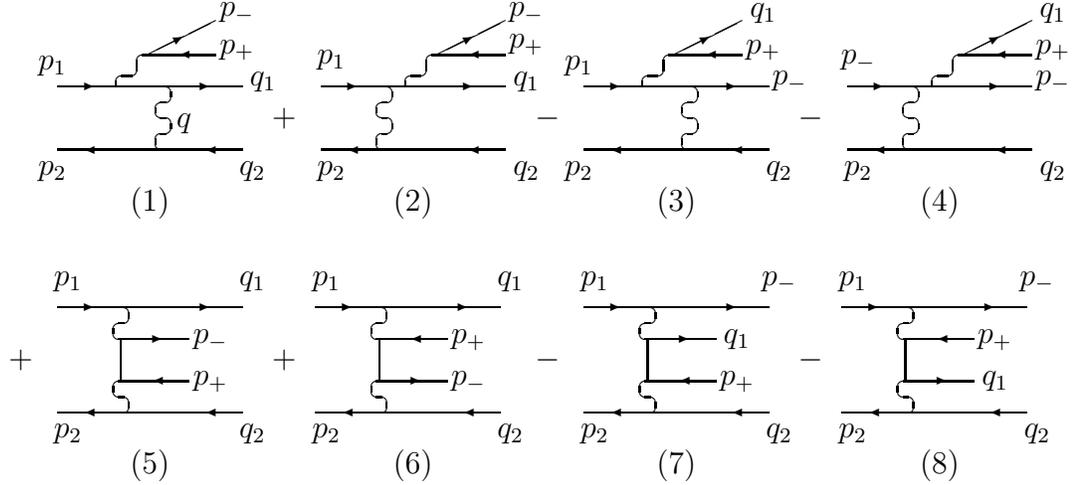
Fig. 1. {\small The Feynman diagram giving logarithmically
enhanced contributions in the kinematical region where the created
pair goes along the electron direction. The signs represent the
Fermi--Dirac statistics of the interchanged fermions.}
\vspace{.3cm}

The result has the factorized form (in agreement with the factorization
theorem [8]):
\begin{eqnarray} \label{eq:10}
\sum\limits_{\mbox{\small spins}} |M|^2 \Big|_{p_+,p_-\parallel p_1}
=\sum\limits_{\mbox{\small spins}} |M_0|^2 \, 2^7 \pi^2 \alpha^2
\frac{I}{m^4}\, ,
\end{eqnarray}
where one of the factors corresponds to the matrix element
in the Born approximation (without pair production):
\begin{eqnarray} \label{eq:11}
&& \sum\limits_{\mbox{spins}} |M_0|^2=2^7\pi^2\alpha^2
\left(\frac{s^4+t^4+u^4}{s^2t^2}\right), \\ \nonumber
&& \qquad s=2p_1p_2x, \qquad t=-Q^2x,
\qquad u=-s-t,
\end{eqnarray}
and the quantity $I$, the collinear factor,
coincides with the expression for $I_a$ obtained in paper [6].
We put it here in terms of our kinematical variables:
\begin{eqnarray} \label{eq:12}
I&=&(1-x_2)^{-2} \left(\frac{A(1-x_2)+Dx_2}{DC}\right)^2
+ (1-x)^{-2} \left(\frac{C(1-x)-Dx_2}{AD}\right)^2 \nonumber \\
&+&\frac{1}{2xAD}\left[ \frac{2(1-x_2)^2-(1-x)^2}{1-x}
+ \frac{x_1x-x_2}{1-x_2} + 3(x_2-x) \right] \nonumber \\
&+& \frac{1}{2xCD}\biggl[ \frac{(1-x_2)^2-2(1-x)^2}{1-x_2}
+ \frac{x-x_1x_2}{1-x} + 3(x_2-x) \biggr]  \nonumber \\
&+& \frac{x_2(x^2+x_2^2)}{2x(1-x_2)(1-x)AC} + \frac{3x}{D^2}
+ \frac{2C}{AD^2} + \frac{2A}{CD^2} + \frac{2(1-x_2)}{xA^2D}  \nonumber \\
&-& \frac{4C}{xA^2D^2} - \frac{4A}{D^2C^2}
+ \frac{1}{DC^2}\left[ \frac{(x_1-x)(1+x_2)}{x(1-x_2)}
- 2\frac{1-x}{x}\right].
\end{eqnarray}
Rearranging the phase volume of the final particles as follows:
\begin{eqnarray} \label{eq:13}
\dd \Gamma&=&\frac{\dd^3q_1\dd^3q_2}{(2\pi)^6 2q_1^02q_2^0}(2\pi)^4
\delta^4(p_1x+p_2-q_1-q_2)\\ \nonumber  &\times&
m^42^{-8}\pi^{-4}x_1x_2\dd x_1\dd x_2
\dd z_1 \dd z_2 \frac{\dd \phi}{2\pi}\, ,
\end{eqnarray}
and integrating over the variables of the created pair, we obtain
(see Appendix A):
\begin{eqnarray} \label{eq:14}
\bar{I}&=&\int\limits_{0}^{2\pi}\!\!\!\;\frac{\dd\phi}{2\pi}
\int\limits_{0}^{z_0}\!\!\!\;\dd z_1 \int\limits_{0}^{z_0}\!\!\!\;\dd z_2 \, I
=\frac{L_0}{2xx_1x_2} \biggl\{D_1\bigl(L_0+2\ln\frac{x_1x_2}{x}\bigr)
\\ \nonumber
&+& D_2\ln\frac{(1-x_2)(1-x)}{xx_2} + D_3 \biggr\},\qquad
L_0=\ln\left(\frac{\varepsilon\theta_0}{m} \right)^2, \\ \nonumber
D_1&=&2xx_1x_2\left(\frac{1}{(1-x)^4}+\frac{1}{(1-x_2)^4} \right)
- \frac{(1-x_2)^2}{(1-x)^2} - \frac{(1-x)^2}{(1-x_2)^2} + 1 \\ \nonumber
&+& \frac{(x+x_2)^2}{2(1-x)(1-x_2)}
+ \frac{3(x_2-x)^2}{2(1-x)(1-x_2)} - \frac{x^2+x_2^2}{(1-x)(1-x_2)} \\
\nonumber
&-& 2xx_2\left(\frac{1}{(1-x)^2}+\frac{1}{(1-x_2)^2} \right)\, , \qquad
D_2=\frac{2(x^2+x_2^2)}{(1-x)(1-x_2)}\, , \\ \nonumber
D_3&=&\frac{2xx_1x_2}{(1-x_2)^2}\left(-\frac{8}{(1-x_2)^2}
+ \frac{(1-x)^2}{xx_1x_2} \right) + \frac{2xx_1x_2}{(1-x)^2}
\biggl[ \frac{x_2}{xx_1}
\\ \nonumber
&+& \frac{2(x_1-x_2)}{xx_1(1-x)} - \frac{8}{(1-x)^2}
+ \frac{1}{xx_1x_2} - \frac{4}{x(1-x)} \biggr] + 6
+ 4x\biggl[\frac{x_2-x_1}{(1-x)^2}
\\ \nonumber
&-& \frac{x_1}{x(1-x)} \biggr] + \frac{4(xx_2-x_1)}{(1-x_2)^2}
- \frac{4(1-x_2)x_1x_2}{(1-x)^3} + \frac{8xx_1x_2^2}{(1-x)^4}
\\ \nonumber
&-& \frac{xx_2^2}{(1-x_2)^4} + \frac{x_2}{(1-x_2)^2} \biggl[4(1-x)
+ \frac{2(x-x_1)(1+x_2)}{1-x_2} \biggr]\, .
\end{eqnarray}
By doing the same also in the case of a pair moving
in the direction of the scattered electron, integrating the obtained
sum over the energy fractions of the pair components, and finally adding
the contribution of the two remaining CK regions (when the pair goes
along the positron direction) we obtain:
\begin{eqnarray} \label{eq:15}
\dd \sigma_{\mbox{\small coll}} &=& \frac{\alpha^4\dd x}{\pi Q_1^2}
\int\limits_{1}^{\rho^2}\!\!\!\;\frac{\dd z}{z^2}L
\biggl\{R_0(x)\left(L+2\ln\frac{\lambda^2}{z}\right)
(1+\Theta) \\ \nonumber
&+& 4R_0(x)\ln x + 2\Theta f(x) + 2f_1(x) \biggr\}\, , \qquad
\lambda=\frac{\theta_0}{\theta_{\mbox{\small min}}},\\ \nonumber
\Theta &\equiv& \Theta(x^2\rho^2-z)=
\left\{ \begin{array}{l}  1,\quad x^2\rho^2 > z, \\ 0,\quad x^2\rho^2 \leq z,
\end{array} \right. \\ \nonumber
R_0(x)&=&\frac{2}{3}\,\frac{(1+x^2)}{1-x} + \frac{(1-x)}{3x}
(4+7x+4x^2) + 2(1+x)\ln x, \\ \nonumber
f(x)&=& - \frac{107}{9} + \frac{136}{9}x - \frac{2}{3}x^2 - \frac{4}{3x}
- \frac{20}{9(1-x)} + \frac{2}{3} \bigl[ - 4x^2 - 5x + 1 \\ \nonumber
&+& \frac{4}{x(1-x)} \bigr] \ln (1-x) + \frac{1}{3} \bigl[
8x^2+5x-7-\frac{13}{1-x} \bigr] \ln x
- \frac{2}{1-x}\ln^2x \\ \nonumber
&+& 4(1+x)\ln x \ln(1-x) - \frac{2(3x^2-1)}{1-x}\mbox{Li}_2(1-x), \\ \nonumber
f_1(x)&=&-x\Re \mbox{e} f(\frac{1}{x})= - \frac{116}{9}
+ \frac{127}{9}x + \frac{4}{3}x^2
+ \frac{2}{3x} - \frac{20}{9(1-x)} + \frac{2}{3} \bigl[ - 4x^2
\\ \nonumber
&-& 5x + 1 + \frac{4}{x(1-x)} \bigr] \ln (1-x)
+ \frac{1}{3} \bigl[8x^2-10x-10+\frac{5}{1-x} \bigr] \ln x
\\ \nonumber
&-& (1+x)\ln^2x + 4(1+x)\ln x \ln(1-x)
- \frac{2(x^2-3)}{1-x}\mbox{Li}_2(1-x), \\ \nonumber
\mbox{Li}_2(x) &\equiv & - \int\limits_{0}^{x}\!\!\!\;\frac{\dd y}{y}\ln(1-y),
\qquad Q_1=\varepsilon\theta_{\mbox{\small min}},
\qquad L=\ln\frac{zQ_1^2}{m^2},
\end{eqnarray}
Some misprints in the expressions for $f(x)$ and
$f_1(x)$ in [4,6] are corrected here.

\section{The Semi-Collinear Kinematics}

We will restrict ourselves again to the case when the created pair
goes close to the electron momentum (the initial or final one).
Analogous considerations can be made in the CM system in the
case when the pair follows the positron momentum.
There are three different semi-collinear
regions, which contribute to the cross-section within
the required accuracy of $O(0.1\%)$. The first region includes the events with
a very small invariant mass of the created pair:
\begin{eqnarray*}
4m^2 \ll (p_++p_-)^2 \ll |q^2|,
\end{eqnarray*}
when the pair escapes the narrow cones (defined by $\theta_0$)
along both the projectile and the scattered electron momentum directions.
We represent this semi-collinear kinematics (SCK) region with the notation
$\vec{p}_+\parallel\vec{p}_-$.
Only the diagrams in fig.~1(1) and fig.~1(2) do contribute to this region
and this is because of the smallness of the virtuality of the pair
producing photon.

The second SCK region includes the events where the invariant mass
of the created positron and the scattered electron is small:
$4m^2 \ll (p_++q_1)^2 \ll |q^2|$, with the restriction that the positron
should escape the narrow cone along the initial electron momentum
direction. We represent it by $\vec{p}_+\parallel\vec{q}_1$ and note that
only two diagrams fig.~1(3) and fig.~1(4), contribute here.

The third SCK region includes the events when the created electron
goes inside the narrow cone along the initial electron momentum
direction but the created positron does not. We represent it by
$\vec{p}_-\parallel\vec{p}_1$. Only the diagrams fig.~1(7) and fig.~1(8)
are relevant here.

The differential cross--section takes the following form:
\begin{eqnarray} \label{eq:16}
\dd \sigma &=& \frac{\alpha^4}{8\pi^4 s^2}
\, \frac{|M|^2}{q^4} \, \frac{\dd x_1 \dd x_2 \dd x}{x_1x_2x}
\dd^2p_+^{\bot} \dd^2p_-^{\bot} \dd^2q_1^{\bot} \dd^2q_2^{\bot}
\delta(1-x_1-x_2-x) \\ \nonumber &\times &
\delta^{(2)}(p_+^{\bot} + p_-^{\bot} + q_1^{\bot} + q_2^{\bot})\, ,
\qquad |M|^2=-L_{\lambda\rho}p_{2\lambda}p_{2\rho},
\end{eqnarray}
where $x_1$ $(x_2)$, $x$ and $p_+^{\bot}$ $(p_-^{\bot})$, $q_1^{\bot}$
are the energy fractions and the perpendicular momenta of the created
positron (electron) and the scattered electron respectively;
$s=(p_1+p_2)^2$ and $q^2=-Q^2=(p_2-q_2)^2=-\varepsilon^2\theta^2$
are the centre-of-mass energy squared and the squared momentum transferred;
the leptonic tensor $L_{\lambda\rho}$ has different forms
for different SCK regions.

\subsection{$\vec{p}_+\parallel\vec{p}_-$ region}

For the region of small $(p_++p_-)^2$ we can use the leptonic
tensor obtained in paper [6]. Keeping only the
relevant terms, we present it in the form:
\begin{eqnarray} \label{eq:17}
\frac{P^4}{8} L_{\lambda\rho}&=&\frac{4P^2q^2}{(1)(2)} \bigl[
- (p_1p_1)_{\lambda\rho} - (q_1q_1)_{\lambda\rho} + (p_1q_1)_{\lambda\rho}
\bigr] \nonumber \\ \nonumber
&-& 4(p_+p_-)_{\lambda\rho}\left(1-\frac{q^2P^2}{(1)(2)}\right)
- \frac{4}{(1)}\bigl[ q^2(p_1q_1)_{\lambda\rho}
- 2(p_1p_+)(q_1p_-)_{\lambda\rho} \\ \nonumber
&-& 2(p_1p_-)(q_1p_1)_{\lambda\rho} \bigr]
- \frac{4}{(2)}\bigl[ P^2(p_1q_1)_{\lambda\rho}
- 2(p_+q_1)(p_1p_-)_{\lambda\rho} \\ \nonumber
&-& 2(p_-q_1)(p_1p_+)_{\lambda\rho} \bigr]
- \frac{32(p_1p_+)(p_1p_-)}{(1)^2}(q_1q_1)_{\lambda\rho}
- \frac{32(q_1p_+)(q_1p_-)}{(2)^2}(p_1p_1)_{\lambda\rho}
\\
&+& \frac{8(p_1q_1)_{\lambda\rho}}{(1)(2)} \bigl[
P^2(p_1q_1) -2(p_1p_+)(p_-q_1) - 2(p_1p_-)(q_1p_+) \bigr],
\end{eqnarray}
where
\begin{eqnarray*}
P&=&p_++p_-, \qquad (aa)_{\lambda\rho}=a_{\lambda}a_{\rho},
\qquad (ab)_{\lambda\rho}=a_{\lambda}b_{\rho}+a_{\rho}b_{\lambda},\\
q&=&p_1-q_1-P,\qquad (1)=(p_1-P)^2-m^2, \qquad (2)=(p_1-q)^2-m^2.
\end{eqnarray*}
After some algebraic transformations the expression for $|M|^2$
entering the cross-section can be put in the form:
\begin{eqnarray} \label{eq:18}
\frac{1}{q^4}|M|^2 &=& - \frac{2s^2}{q^4P^4}\biggl\{-\frac{4P^2q^2}{(1)(2)}
\bigl[(1-x_1)^2+(1-x_2)^2\bigr] \nonumber \\
&+& \frac{128}{(1)^2(2)^2}\bigl[(q_1p)\,(p_+p_1) - x(p_1p)\,(q_1p_+)\bigr]^2
\biggr\},
\end{eqnarray}
where $p=p_- -x_2p_+/x_1$, $\ (q_2^{\bot})^2=-q^2$.
In the considered region we can use the relations
\begin{eqnarray} \label{eq:19}
(1)=-\frac{1-x}{x_1}\, 2(p_1p_+), \qquad (2)=\frac{1-x}{x_1}\, 2(q_1p_+).
\end{eqnarray}

It is useful to represent all invariants in terms of the Sudakov
variables (energy fractions and perpendicular momenta), namely
\begin{eqnarray} \label{eq:20}
&& q_1^2=\frac{1}{x_1x_2}((p^{\bot})^2+m^2(1-x)^2), \quad
2(q_1p_+)=\frac{1}{xx_1}(xp_+^{\bot} - x_1q_1^{\bot})^2,
\\ \nonumber
&& 2(p_1p_+)=\frac{1}{x_1}(p_+^{\bot})^2, \quad
2(p_1p)=\frac{2}{x_1^2}p^{\bot}p_+^{\bot}, \quad
2(q_1p)=\frac{2}{x_1^2}(p^{\bot}\,[xp_+^{\bot} - x_1q_1^{\bot}]),
\\ \nonumber
&& p^{\bot}=x_1p_-^{\bot} - x_2p_+^{\bot} \, .
\end{eqnarray}

The large logarithm appears in the cross-section after the integration
over $p^{\bot}$. In order to carry out this integration we can use the
relation
\begin{eqnarray} \label{eq:21}
\delta^{(2)}\dd^2p_+^{\bot}\dd^2p_-^{\bot}=\frac{1}{(1-x)^2}\dd^2p^{\bot},
\end{eqnarray}
which is valid in the region $\vec{p}_+\parallel\vec{p}_-$.
After the integration we derive the contribution of the first
SCK region to the cross-section:
\begin{eqnarray} \label{eq:22}
\dd \sigma_{\vec{p}_+\parallel\vec{p}_-} &=& \frac{\alpha^4}{\pi} L
\;\dd x\; \dd x_2 \;\frac{\dd (q_2^{\bot})^2}{(q_2^{\bot})^2} \cdot
\frac{\dd (q_1^{\bot})^2}{(q_1^{\bot}+q_2^{\bot})^2}
\\ \nonumber &\times&
\frac{\dd \phi}{2\pi} \cdot \frac{1}{(q_1^{\bot}+xq_2^{\bot})^2}
\biggl[(1-x_1)^2+(1-x_2)^2-\frac{4xx_1x_2}{(1-x)^2} \biggr],
\end{eqnarray}
where $\phi$ is the angle between the two dimensional vectors
$q_1^{\bot}$ and $q_2^{\bot}$.

At this stage it is necessary to use the restrictions on the two
dimensional momenta $q_1^{\bot}$ and $q_2^{\bot}$. They appear when the
contribution of the CK region (which in this case represents the narrow
cones with the opening angle $\theta_0 $ along the momentum directions
of both initial and scattered electron) is excluded.

%-----------------fig.2-----------------------------------
%\begin{figure}
\vspace{.7cm}
\unitlength=1.0mm
\special{em:linewidth 0.5pt}
\linethickness{0.5pt}
\begin{picture}(130.00,65.00)
\put(40.00,20.00){\vector(-3,-1){18.00}}
\put(22.00,14.00){\line(-3,-1){18.00}}
\put(40.00,20.00){\oval(6.00,6.00)}
\put(40.00,20.00){\vector(1,3){6.00}}
\put(46.00,38.00){\line(1,3){6.00}}
\put(52.00,56.00){\oval(6.00,6.00)}
\put(40.00,20.00){\vector(2,-1){24.00}}
\put(4.00,8.00){\vector(1,1){24.00}}
\put(28.00,32.00){\line(1,1){24.00}}
\put(25.00,34.00){\makebox(0,0)[cc]{$r^{\bot}$}}
\put(69.00,8.00){\makebox(0,0)[cc]{$q_2^{\bot}/\varepsilon$}}
\put(25.00,9.00){\makebox(0,0)[cc]{$p_+^{\bot}/\varepsilon_+$}}
\put(51.00,36.00){\makebox(0,0)[cc]{$q_1^{\bot}/\varepsilon_2$}}
\end{picture}
%\caption{The kinematics of an event in the perpendicular angular plane
%corresponding to the SCK region $\vec p_+ \parallel \vec p_-$.}
%\end{figure}

\nopagebreak
\vspace{.1cm}
\noindent
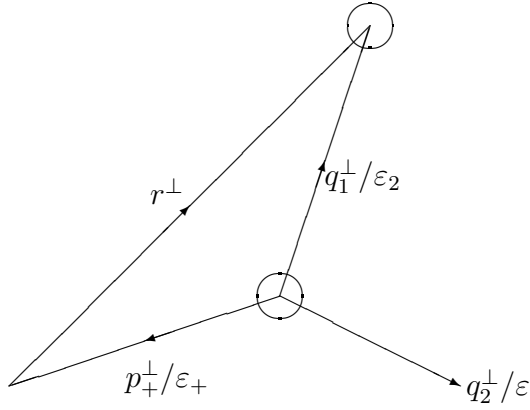
Fig.~2. {\small The kinematics of an event in the angular perpendicular
plane corresponding to the SCK region $\vec{p}_+ \parallel \vec{p}_-$.}
\vspace{.3cm}

The kinematics of the events corresponding to the region
$\vec{p}_+ \parallel \vec{p}_-$ in the perpendicular plane is shown in
fig.~2. The circles radius $\theta_0 $ represent in the figure
the forbidden
collinear regions. The elimination of those regions gives the following
restrictions:
\begin{equation} \label{eq:23}
|\frac{p_+^{\bot}}{\varepsilon_+}| > \theta_0, \qquad
|r^{\bot}|=|\frac{p_+^{\bot}}{\varepsilon_+}
-\frac{q_1^{\bot}}{\varepsilon_2}| > \theta_0,
\end{equation}
where $\varepsilon_+ $ and $ \varepsilon_2 $ are the energies of the
created positron and of the scattered electron respectively. In order
to exclude $p_+^{\bot}$ from the above equation we use the
conservation of the perpendicular momentum in the region
$p_+ \parallel \vec{p}_-$:
\begin{eqnarray} \label{eq:24}
q_1^{\bot}+q_2^{\bot}+\frac{1-x}{x_1}p_+^{\bot}=0.
\end{eqnarray}

It is useful to introduce the dimensionless variables
$z_{1,2}=(q_{1,2}^{\bot})^2/(\varepsilon \theta_{\mbox{min}})^2$,
where $\theta_{\mbox{\small min}}$ is the minimal angle at which the scattered
particles (electron and positron) are recorded by the detector.
Here we consider only the symmetrical circular detectors.
The conditions (\ref{eq:23}) can be rewritten as follows:

\begin{equation} \label{eq:25}
\left\{ \begin{array}{l}
1 > \cos \phi > - 1 + \frac{\lambda^2(1-x)^2-(\sqrt{z_1}-\sqrt{z_2})^2}
{2\sqrt{z_1z_2}},\qquad
|\sqrt{z_1}-\sqrt{z_2}| < \lambda (1-x), \\
1 > \cos \phi > - 1 ,\qquad |\sqrt{z_1}-\sqrt{z_2}| > \lambda (1-x),
\qquad \lambda=\theta_0/\theta_{\mbox{\small min}}, \end{array} \right.
\end{equation}

\begin{equation} \label{eq:26}
\left\{ \begin{array}{l}
1 > \cos \phi > - 1
+ \frac{\lambda^2x^2(1-x)^2-(\sqrt{z_1}-x\sqrt{z_2})^2}
{2x\sqrt{z_1z_2}},\qquad
|\sqrt{z_1}-x\sqrt{z_2}| < \lambda x(1-x), \\ \label{eq:27}
1 > \cos \phi > - 1 ,\qquad
|\sqrt{z_1}-x\sqrt{z_2}| > \lambda x(1-x). \end{array} \right.
\end{equation}

The restrictions in eq.(\ref{eq:25})[(\ref{eq:26})] exclude the phase space,
corresponding to the narrow cone along the direction of the initial [scattered]
electron.

The conditions of the LEP~I experiment are:
\begin{eqnarray}
\theta_0 \gg \frac{m}{\varepsilon} \approx 10^{-5} \quad
\mbox{and} \quad \theta_{\mbox{\small min}} \sim 10^{-2}.
\end{eqnarray}
This is the reason for considering $\lambda \ll 1$.
The procedure
of integration of the differential cross-section over regions
(\ref{eq:25}) and (\ref{eq:26}) is described in detail in Appendix B.
Here we give the contribution of the SCK region
$\vec{p}_+ \parallel \vec{p}_-$ to the cross-section provided that only
the scattered electrons with the energy fraction $x$ exceeding $x_c$
could be recorded:
\begin{eqnarray} \label{eq:28}
\sigma_{\vec{p}_+ \parallel \vec{p}_-} &=& \frac{\alpha^4}{\pi Q_1^2}
{\cal L} \int\limits_{1}^{\rho^2}\!\!\!\; \frac{\dd z}{z^2}
\int\limits_{x_c}^{1-\Delta}\!\!\!\;\dd x \int\limits_{0}^{1-x}\!\!\!\; \dd x_2
\biggl[ \frac{(1-x_1)^2+(1-x_2)^2}{(1-x)^2}
\nonumber \\ \nonumber
&-& \frac{4xx_1x_2}{(1-x)^2} \biggr]
\biggl\{ (1+\Theta)\ln \frac{z}{\lambda^2}
+ \Theta \ln \frac{(x^2\rho^2-z)^2}{x^2(x\rho^2-z)^2}
\\
&+& \ln \left| \frac{(z-x^2)(\rho^2-z)(z-1)}{(z-x)^2(z-x^2\rho^2)} \right|
\biggr\}, \qquad {\cal L}=\ln \frac{\varepsilon^2\theta_{\mbox{\small
min}}}{m^2}\, ,
\end{eqnarray}
where $Q_1^2=\varepsilon^2\theta_{\mbox{\small min}}^2$,
$\rho=\theta_{\mbox{\small max}}/\theta_{\mbox{\small min}}$
($\theta_{\mbox{\small max}}$ is the maximal angle of the final particle
registration),
$\Theta \equiv \Theta(x^2\rho^2-z)$, $z\equiv z_2$. The auxiliary parameter
$\Delta$ entering eq.~(\ref{eq:28}) defines the minimal energy of the created
hard pair: $2m/\varepsilon \ll \Delta \ll 1$. Note that we replaced $L$
by ${\cal L}$ because we do make no difference between them at the single
logarithmic level.

\subsection{$\vec{p}_+\parallel\vec{q}_1$ region}

As was already mentioned, in the SCK region $\vec{p}_+\parallel\vec{q}_1$
only diagrams fig.~1(3) and fig.~1(4) contribute. The leptonic tensor
could in this case be derived from eq.~(\ref{eq:17}) by the substitution
$p_- \leftrightarrow q_1$, and the squared matrix element could
be written as
\begin{eqnarray} \label{eq:29}
|M|^2_{\vec{p}_+\parallel\vec{q}_1} &=&
- \frac{4s^2}{{q'}_1\,\!\!^2 \vec{q}_2\,\!\!^2} \cdot \frac{1}{(1')(2)}
\biggl\{ (1-x_1)^2+(1-x_2)^2
\nonumber \\
&+& \frac{32}{{q'}_1\,\!\!^2 \vec{q}_2\,\!\!^2}
\cdot \frac{1}{(1')(2)} \bigl[ (p_1p_+)(p_-p')
-x_2(p_-p_+)(p_1p')\bigr]^2  \biggr\},
\end{eqnarray}
where
\begin{eqnarray*}
p'&=&q_1-p_+x/x_1,\qquad {q'}_1\,\!\!^2=(q_1+p_+)^2, \\
(2)&=&2(p_+p_-)(1-x_2)/x_1,\qquad (1')=-2(p_1p_+)(1-x_2)/x_1.
\end{eqnarray*}

The integration of the matrix element over $(p_1^{\bot})^2$ and
$(p_-^{\bot})^2$ could be carried out analogously to the previous case,
and the contribution of the $\vec{p}_+\parallel\vec{q}_1$ region
could be presented in the following form:
\begin{eqnarray} \label{eq:30}
\dd \sigma_{\vec{p}_+\parallel\vec{q}_1} &=& \frac{\alpha^4}{\pi} L
\;\dd x\; \dd x_2\; \frac{\dd (q_2^{\bot})^2}{(q_2^{\bot})^2} \cdot
\frac{\dd (q_1^{\bot})^2}{(q_1^{\bot})^2}
\nonumber  \\ \nonumber &\times&
\frac{\dd \phi}{2\pi} \cdot \frac{1}{(q_1^{\bot}+xq_2^{\bot})^2}
\cdot \frac{x^2}{(1-x_2)^2}
\biggl[\frac{(1-x)^2+(1-x_1)^2}{(1-x_2)^2}
-\frac{4xx_1x_2}{(1-x_2)^2} \biggr].\\
\end{eqnarray}

The restriction on the phase space, coming from the exclusion
of the collinear region when the created pair flies inside the
narrow cone along the scattered electron, leads to the relation:
\begin{eqnarray} \label{eq:31}
\left| \frac{p_-^{\bot}}{\varepsilon_-}
- \frac{q_1^{\bot}}{\varepsilon_2} \right| > \theta_0.
\end{eqnarray}
In eq.~(\ref{eq:31}) we have to exclude $p_-^{\bot}$ using the
conservation of the perpendicular momentum in the case
under consideration: $p_-^{\bot} + q_2^{\bot} + q_1^{\bot}(1-x_2)/x=0$.
In terms of the dimensionless variables $z_1,\,z_2$ and the angle $\phi$,
eq.~(\ref{eq:31}) could be rewritten as
\begin{equation}\label{eq:32}
\left\{ \begin{array}{l}
1 > \cos \phi > - 1
+ \frac{\lambda^2x^2x_2^2-(\sqrt{z_1}-x\sqrt{z_2})^2}
{2x\sqrt{z_1z_2}},\qquad
|\sqrt{z_1}-x\sqrt{z_2}| < \lambda x x_2, \\
1 > \cos \phi > - 1 ,\qquad
|\sqrt{z_1}-x\sqrt{z_2}| > \lambda x x_2. \end{array} \right.
\end{equation}

The integration of the differential cross-section (\ref{eq:30})
over the region defined in eq.~(\ref{eq:32}) leads to the following
result for the contribution of the $\vec{p}_+\parallel\vec{q}_1$
SCK region:
\begin{eqnarray} \label{eq:33}
\sigma_{\vec{p}_+ \parallel \vec{q}_1} &=& \frac{\alpha^4}{\pi Q_1^2}
{\cal L} \int\limits_{1}^{\rho^2}\!\!\!\; \frac{\dd z}{z^2}
\int\limits_{x_c}^{1-\Delta}\!\!\! \dd x \int\limits_{0}^{1-x}\!\!\! \dd x_2
\biggl[ \frac{(1-x)^2+(1-x_1)^2}{(1-x_2)^2}
\nonumber \\
&-& \frac{4xx_1x_2}{(1-x_2)^4} \biggr]
\biggl\{ \ln \frac{z}{\lambda^2}
+ \ln \frac{(\rho^2-z)(z-1)}{x_2^2\rho^2} \biggr\}.
\end{eqnarray}

\subsection{$\vec{p}_-\parallel\vec{p}_1$ region}

In the SCK region $\vec{p}_-\parallel\vec{p}_1$,
only diagrams fig.~1(7) and fig.~1(8) contribute to the cross-section
within the required accuracy. In this case, the leptonic tensor
could be derived from eq.~(\ref{eq:17}) by the substitution
$p_1 \leftrightarrow -p_+$, and the squared matrix element has the form:
\begin{eqnarray} \label{eq:34}
|M|^2_{\vec{p}_-\parallel\vec{p}_1} &=&
- \frac{4s^2}{{q'}_2\,\!\!^2 \vec{q}_2\,\!\!^2} \cdot \frac{1}{(1)(2')}
\biggl\{ (1-x)^2+(1-x_1)^2 \nonumber
\\
&+& \frac{32}{{q'}_2\,\!\!^2 \vec{q}_2\,\!\!^2}
\cdot \frac{1}{(1)(2')} \bigl[ x_1(p_1\tilde{p})(p_1p_+)
+x(p_+\tilde{p})(q_1p_1)\bigr]^2  \biggr\},
\end{eqnarray}
where
\begin{eqnarray*}
\tilde{p}&=&p_- - x_2p_1,\qquad {q'}_2\,\!\!^2=(p_1-p_-)^2, \\
(2')&=&-2(p_1q_1)(1-x_2),\qquad (1)=-2(p_1p_+)(1-x_2).
\end{eqnarray*}

The integration of the matrix element over $(p_+^{\bot})^2$ and
$(p_-^{\bot})^2$ leads to the differential cross-section
\begin{eqnarray} \label{eq:35}
\dd \sigma_{\vec{p}_-\parallel\vec{p}_1} &=& \frac{\alpha}{4\pi} L
\;\dd x \;\dd x_2 \;\frac{\dd (q_2^{\bot})^2}{(q_2^{\bot})^2} \cdot
\frac{\dd (q_1^{\bot})^2}{(q_1^{\bot})^2} \nonumber
\\ &\times&
\frac{\dd \phi}{2\pi} \cdot \frac{1}{(q_1^{\bot}+q_2^{\bot})^2}
\biggl[\frac{(1-x)^2+(1-x_1)^2}{(1-x_2)^2}-\frac{4xx_1x_2}{(1-x_2)^4} \biggr].
\end{eqnarray}

The restriction due to the exclusion
of the collinear region when the created pair flies inside a
narrow cone along the initial electron has the form:
\begin{eqnarray} \label{eq:36}
\frac{|p_+^{\bot}|}{\varepsilon_1} > \theta_0, \qquad
p_+^{\bot} + q_1^{\bot} + q_2^{\bot}=0,
\end{eqnarray}
or
\begin{equation}\label{eq:37}
\left\{ \begin{array}{l}
1 > \cos \phi > - 1
+ \frac{\lambda^2 x_1^2-(\sqrt{z_1}-\sqrt{z_2})^2}
{2\sqrt{z_1z_2}},\qquad
|\sqrt{z_1}-\sqrt{z_2}| < \lambda x_1, \\
1 > \cos \phi > - 1 ,\qquad
|\sqrt{z_1}-\sqrt{z_2}| > \lambda x_1.
\end{array} \right.
\end{equation}

The integration of the differential cross-section (\ref{eq:35})
over the region defined in eq.~(\ref{eq:37}) leads to
\begin{eqnarray} \label{eq:38}
\sigma_{\vec{p}_- \parallel \vec{p}_1} &=& \frac{\alpha^4}{\pi Q_1^2}
{\cal L} \int\limits_{1}^{\rho^2}\!\!\!\; \frac{\dd z}{z^2}
\int\limits_{x_c}^{1-\Delta}\!\!\!\dd x \int\limits_{0}^{1-x}\!\!\! \dd x_2
\biggl[ \frac{(1-x)^2+(1-x_1)^2}{(1-x_2)^2}
 \nonumber \\
&-& \frac{4xx_1x_2}{(1-x_2)^2} \biggr]
\biggl\{ \Theta \ln \frac{z}{\lambda^2}
+ \Theta \ln \frac{(x^2\rho^2-z)^2}{x_1^2x^4\rho^4}
+ \ln \left| \frac{\rho^2(z-x^2)}{z-x^2\rho^2} \right| \biggr\}.
\end{eqnarray}

The total contribution of the semi-collinear kinematics
to the cross-section is the sum of eqs.~(\ref{eq:28}), (\ref{eq:33}), and
(\ref{eq:38}):
\begin{eqnarray} \label{eq:39}
\sigma_{\mbox{\small s-coll}}=\sigma_{\vec{p}_+ \parallel \vec{p}_-}
+ \sigma_{\vec{p}_+ \parallel \vec{q}_1}
+ \sigma_{\vec{p}_- \parallel \vec{p}_1}.
\end{eqnarray}

\section{The Total Contribution Due to the Real
and Virtual Pair Production}

In order to obtain finite expression for the cross-section
we have to add to eq.~(\ref{eq:39}) the contribution of the
collinear kinematics region (eq.~(\ref{eq:15}))
as well as the ones due to the production
of virtual and soft pairs. Taking into account the leading and
next-to-leading terms we can write the full hard pair contribution
in the following form:
\begin{eqnarray} \label{eq:40}
\sigma_{\mbox{\small hard}}&=&\frac{\alpha^4}{\pi Q_1^2}\int
\limits_{1}^{\rho^2}\!\!\!
\;\frac{\dd z}{z^2} \int\limits_{x_c}^{1-\Delta}\!\!\! \dd x
\biggl\{ L^2R(x) + {\cal L}[\Theta f(x)+f_1(x)]
\\ \nonumber
&+& {\cal L}\int\limits_{0}^{1-x}\!\!\!\dd x_2 \biggl[
\biggl( \Theta \ln \frac{(x^2\rho^2-z)^2}{x^2}
+ \ln \left| \frac{(z-x^2)(\rho^2-z)(z-1)x^2}{z-x^2\rho^2} \right|
\biggr) \varphi \\ \nonumber
&-& \bigl(\Theta \ln (x\rho^2-z)^2 + \ln (z-x)^2 \bigr)\varphi(x,x_2)
\\ \nonumber
&-& \bigl(\Theta \ln (x_1^2x^2\rho^4) + \ln x_2^2 \bigr)\varphi(x_2,x)
\biggr] \biggr\},
\qquad L=\ln \frac{Q_1^2z}{m^2},\quad {\cal L}=\ln \frac{Q_1^2}{m^2},
\end{eqnarray}
where
\begin{eqnarray*}
\varphi &=&\varphi(x,x_2)+\varphi(x_2,x), \\
\varphi(x_2,x)&=&\frac{(1-x)^2+(x+x_2)^2}{(1-x_2)^2}
- \frac{4xx_2(1-x-x_2)}{(1-x_2)^4}, \\
R(x)&=&\frac{1}{3}\cdot \frac{1+x^2}{1-x} + \frac{1-x}{6x}(4+7x+4x^2)
+ (1+x)\ln x.
\end{eqnarray*}
Integrating over $x_2$ in the right-hand side of eq.~(\ref{eq:40})
we obtain the final expression for the cross-section of hard
pair production at small angle electron--positron scattering:
\begin{eqnarray} \label{eq:41}
\sigma_{\mbox{\small hard}}&=&\frac{\alpha^4}{\pi
Q_1^2}\int\limits_{1}^{\rho^2}\!\!\!
\;\frac{\dd z}{z^2} \int\limits_{x_c}^{1-\Delta}\!\!\! \dd x
\biggl\{ L^2(1+\Theta)R(x) + {\cal L}[\Theta F_1(x)+F_2(x)] \biggr\},
\nonumber \\ \nonumber
F_1(x)&=&d(x)+C_1(x),\qquad F_2(x)=d(x)+C_2(x),
\\ \nonumber
d(x)&=&\frac{1}{1-x}\biggl(\frac{8}{3}\ln(1-x)-\frac{20}{9}\biggr),
\\
C_1(x)&=& - \frac{113}{9} + \frac{142}{9}x - \frac{2}{3}x^2
- \frac{4}{3x} - \frac{4}{3}(1+x)\ln(1-x)
\\ \nonumber
&+& \frac{2}{3}\cdot\frac{1+x^2}{1-x} \biggl[
\ln\frac{(x^2\rho^2-z)^2}{(x\rho^2-z)^2} - 3\mbox{Li}_2(1-x) \biggr]
+ \bigl(8x^2+3x-9-\frac{8}{x}
\\ \nonumber
&-&\frac{7}{1-x} \bigr) \ln x + \frac{2(5x^2-6)}{1-x}\ln^2x
+ \beta(x)\ln\frac{(x^2\rho^2-z)^2}{\rho^4},
\\ \nonumber
C_2(x)&=& - \frac{122}{9} + \frac{133}{9}x + \frac{4}{3}x^2
+ \frac{2}{3x} - \frac{4}{3}(1+x)\ln(1-x)
\\ \nonumber
&+& \frac{2}{3}\cdot\frac{1+x^2}{1-x} \biggl[
\ln \left| \frac{(z-x^2)(\rho^2-z)(z-1)}{(x^2\rho^2-z)(z-x)^2} \right|
+ 3\mbox{Li}_2(1-x) \biggr]
\\ \nonumber
&+& \frac{1}{3}\bigl(-8x^2-32x-20+\frac{13}{1-x}+\frac{8}{x} \bigr) \ln x
+ 3(1+x)\ln^2x
\\ \nonumber
&+& \beta(x)\ln\left|\frac{(z-x^2)(\rho^2-z)(z-1)}{x^2\rho^2-z}\right|,
\qquad \beta = 2R(x) - \frac{2}{3}\cdot\frac{1+x^2}{1-x}.
\end{eqnarray}

Formula (\ref{eq:41}) describes the small angle high energy cross-section
of process (\ref{eq:1}) in the case where the created hard pair flies
along the direction of the initial electron three-momentum, and we now have
to double $\sigma_H$  to take into account the production of a hard pair
flying along the direction of the initial positron beam.

In order to pick out the dependence on the parameter $\Delta$ in
$\sigma_H$ we will use the following relation:
\begin{eqnarray} \label{eq:42}
&& \int\limits_{1}^{\rho^2}\!\!\!\;\dd z\int\limits_{x_c}^{1-\Delta}\!\!\!
\dd x\;\Theta(x^2\rho^2-z)
=\int\limits_{1}^{\rho^2}\!\!\!\;\dd z\biggl[
\int\limits_{x_c}^{1-\Delta}\!\!\!
\dd x - \int\limits_{x_c}^{1}\!\!\!\;\dd x \;\bar{\Theta} \biggr],
\\ \nonumber && \qquad
\bar{\Theta}=1-\Theta(x^2\rho^2-z).
\end{eqnarray}
Therefore
\begin{eqnarray} \label{eq:43}
\int\limits_{1}^{\rho^2}\!\!\!\;\dd z\int\limits_{x_c}^{1-\Delta}\!\!\!
\Theta \frac{\dd x}{1-x}&=&\int\limits_{1}^{\rho^2}\!\!\!\;\dd z
\biggl[\ln\frac{1-x_c}{\Delta}-\int\limits_{x_c}^{1}\!\!\!\frac{\dd x}{1-x}
\bar{\Theta} \biggr], \\   \label{eq:44}
\int\limits_{1}^{\rho^2}\!\!\!\;\dd z\int\limits_{x_c}^{1-\Delta}\!\!\!\dd x
\Theta \frac{\ln(1-x)}{1-x}&=&\int\limits_{1}^{\rho^2}\!\!\!\;\dd z
\biggl[\frac{1}{2}\ln^2(1-x_c)-\frac{1}{2}\ln^2\Delta]\\ \nonumber
-\int\limits_{x_c}^{1}\!\!\!\;\dd x\frac{\ln(1-x)}{1-x} \bar{\Theta} .
\end{eqnarray}
The contribution to the cross-section of the small-angle Bhabha scattering
connected with real soft (with an energy less than
$\Delta \cdot \varepsilon$) and virtual pair production is defined [2]
by the formula:
\begin{eqnarray} \label{eq:45}
\sigma_{\mbox{\small soft+virt}}&=&\frac{4\alpha^4}{\pi Q_1^2}
\int\limits_{1}^{\rho^2}\!\!\!\;\frac{\dd z}{z^2} \biggl\{L^2
\biggl(\frac{2}{3}\ln\Delta + \frac{1}{2}\biggr)
+ {\cal L}\biggl(-\frac{17}{6}+\frac{4}{3}\ln^2\Delta
\\ \nonumber
&-& \frac{20}{9} \ln \Delta - \frac{4}{3} \zeta_2\biggr) \biggr\}.
\end{eqnarray}
Using eqs.~(\ref{eq:43}) and (\ref{eq:44}) one can check
that the auxiliary parameter $\Delta$ is cancelled in the sum
$\sigma_{\mbox{\small tot}}=2\sigma_{\mbox{\small hard}}
+\sigma_{\mbox{\small soft+virt}}$,
and we can write the total
contribution $\sigma_{\mbox{\small tot}}$ as follows:
\begin{eqnarray} \label{eq:46}
&& \sigma_{\mbox{\small tot}}=\frac{2\alpha^4}{\pi Q_1^2}
\int\limits_{1}^{\rho^2}\!\!\!
\;\frac{\dd z}{z^2} \biggl\{ L^2\bigl(1+\frac{4}{3}\ln(1-x_c)
-\frac{2}{3}\int\limits_{x_c}^{1}\!\!\!\frac{\dd x}{1-x}\bar{\Theta}\bigr)
+ {\cal L}\biggl[-\frac{17}{3}
\\ \nonumber && \quad
- \frac{8}{3}\zeta_2
-\frac{40}{9}\ln(1-x_c)+\frac{8}{3}\ln^2(1-x_c)
+ \int\limits_{x_c}^{1}\!\!\!\;\frac{\dd x}{1-x}\bar{\Theta}\cdot
\bigl(\frac{20}{9}-\frac{8}{3}\ln(1-x)\bigr) \biggr]
\\ \nonumber && \quad
+ \int\limits_{x_c}^{1}\!\!\!\;\dd x\bigl[ L^2(1+\Theta)\bar{R}(x)
+ {\cal L}(\Theta C_1(x) + C_2(x)) \bigr] \biggr\},
\quad \bar{R}(x)=R(x)-\frac{2}{3(1-x)}.
\end{eqnarray}

The right-hand side of eq.~(\ref{eq:46}) is the master expression for
the small-angle Bhabha scattering cross-section connected with the
pair production. It is finite and could be used for numerical
estimates. Note that the leading term is described by the
electron structure function $D_e^{\bar{e}}(x)$, which represents
the probability to find a positron inside an electron with
virtuality $Q^2$ provided that the electron loses the energy
part $(1-x)$ [9].

In table~1 we present the ratio of the RC contribution due
to the pair production $\sigma_{\mbox{\small tot}}$~(\ref{eq:46})
to the normalization cross--section $\sigma_0$,
\begin{eqnarray} \label{eq:47}
\sigma_0=\frac{4\pi\alpha^2}{\varepsilon^2\theta_{\mbox{\small min}}^2}\, .
\end{eqnarray}

In table~2 we illustrate the comparison between the non-leading
contribution (containing ${\cal L}^1=\ln Q_1^2/m^2$) and
the total one (containing ${\cal L}^2$ and ${\cal L}^1$).

\vspace{.3cm}
%\begin{table}
%\caption{
%The ratio $S=\sigma_{\mbox{\small tot}}/\sigma_{0}$ in per cent,
%as a function of $x_c$, for NN $(\rho=1.74$,
%$\theta_{\mbox{\small min}}=1.61$~rad) and WW $(\rho=2.10$,
%$\theta_{\mbox{min}}=1.50$~rad) counters,
%$\sqrt s=2\varepsilon=M_Z=91.187$ GeV.}
\noindent
Table 1. {\small The ratio $S=\sigma_{\mbox{tot}}/\sigma_{0}$ in percents,
as a function of $x_c$, for NN $(\rho=1.74$,
$\theta_{\mbox{min}}=1.61$~rad) and WW $(\rho=2.10$,
$\theta_{\mbox{min}}=1.50$~rad) counters,
$\sqrt s=2\varepsilon=M_Z=91.187$ GeV.} \\[.3cm]
\begin{tabular}{|c|c|c|c|c|c|c|c|}
\hline
$x_c$         &  0.2  &  0.3  &  0.4  &  0.5  &  0.6  &  0.7  &  0.8  \\
\hline
$S_{NN},\ \%$ &-0.018 &-0.022 &-0.026 &-0.029 &-0.033 &-0.038 &-0.046 \\
\hline
$S_{WW}$,\ \% &-0.013 &-0.019 &-0.024 &-0.029 &-0.035 &-0.042 &-0.052 \\
\hline
\end{tabular}
%\end{table}

\vspace{.3cm}

\vspace{.3cm}
%\begin{table}
%\caption{ Values of $R_{NN}$ and $R_{WW}$
%as functions of $x_c$, where $R$ represents the ratio of
%the non--leading contribution in eq.~(\ref{eq:46}) in
%respect to the total one, for NN and WW counters.}
\noindent
Table 2. {\small Values of $R_{NN}$ and $R_{WW}$
as functions of $x_c$, where $R$ represents the ratio of
the non--leading contribution in eq.~(\ref{eq:46}) with
respect to the total one, for NN and WW counters.} \\[.3cm]
\begin{tabular}{|c|c|c|c|c|c|c|c|}
\hline
$x_c$    &  0.2  &  0.3  &  0.4  &  0.5  &  0.6  &  0.7  &  0.8  \\
\hline
$R_{NN}$ & 0.036 &-0.122 &-0.194 &-0.238 &-0.268 &-0.335 &-0.465 \\
\hline
$R_{WW}$ & 0.179 &-0.021 &-0.088 &-0.120 &-0.179 &-0.271 &-0.415 \\
\hline
\end{tabular}
%\end{table}

\vspace{.3cm}

\section{Conclusions}

The result derived in this paper combined with those
derived earlier in [4,5,6] thus give the full
analytical description of the small angle electron--positron
scattering cross-section at LEP~I energies with one
and two photon radiation as well as pair production.
The description takes into account the leading and next-to-leading
logarithmic approximations and gives the possibility to describe the
cross-section with an accuracy not worse than $0.1\%$ provided that
the scattered electron and positron are recorded by symmetrical
circular detectors. By using the above derivation it is possible to
carry out the calculations also for non-symmetrical detectors.

Numerical calculations of the virtual and real pair production
RC contributions show their compensation at the level of $10^{-3}\%$
for the given angular apertures and $x_c$ range.
Table~2 shows that the next-to-leading contribution
is comparable with the leading one. Their ratio is sensitive
to $x_c$ and angle ranges.

We note that, in a realistic case, one has to take into
account the fact that detectors cannot distinguish a single particle
event from the one when two or more particles hit the same point of
the detector simultaneously. In that case the obtained results could be
easily changed: starting from the presented differential cross-sections
the integration must be done by imposing experimental restrictions.

We want also to emphasize that the method and some of the
derived results could be used for calculating radiative
corrections to deep inelastic scattering as well as
for cross-sections of some normalization processes at HERA.
We hope to consider these questions in a future publication.

\subsection*{Acknowledgements}
We are grateful to V.A.~Fadin and L.N.~Lipatov for fruitful
discussions and criticism. This work is supported in part
by INTAS grant 93--1867.

\section*{Appendix A}

\setcounter{equation}{0}
\renewcommand{\theequation}{A.\arabic{equation}}

We give here a list of the relevant integrals for the collinear kinematical
region, calculated within the logarithmic accuracy. The definitions
of eq.~(\ref{eq:9}) are used. We imply, in the left-hand side of the
relations below the general operation:
\begin{eqnarray} \label{eq:a1}
\langle (\dots) \rangle \equiv \int\limits_{0}^{z_0}\!\!\!\; \dd z_1
\int\limits_{0}^{z_0}\!\!\!\; \dd z_2 \int\limits_{0}^{2\pi}\!\!\!\;
\frac{\dd \phi}{2\pi}\, (\dots)\, ,
\end{eqnarray}
and suggest $z_0=(\varepsilon \theta_0/m)^2 \gg 1$, $\ L_0=\ln z_0 \gg 1$.
The details of the calculations can be found in the Appendix of paper [5].
The results are:
\begin{eqnarray} \label{eq:a2}
&& \langle \left(\frac{x_2D+(1-x_2)A}{DC} \right)^2 \rangle =
\frac{L_0}{(1-x_2)^2} \biggl\{ L_0 + 2\ln\frac{x_1x_2}{x} - 8 \\ \nonumber
&& \qquad + \frac{(1-x)^2(1-x_2)^2}{xx_1x_2} \biggr\}, \qquad
\langle \frac{1}{DC} \rangle = \frac{L_0}{x_1x_2(1-x_2)}
\bigl[\frac{1}{2}L_0 + \ln\frac{x_1x_2}{x} \bigr]\, , \\ \nonumber
&& \langle \left(\frac{x_2A_1-x_1A_2}{AD} \right)^2 \rangle =
\frac{L_0}{(1-x)^2} \biggl\{ L_0 + 2\ln\frac{x_1x_2}{x} - 8
+ \frac{(1-x)^2}{xx_1x_2} - \frac{4(1-x)}{x} \biggr\}\, , \\ \nonumber
&& \langle \frac{x_1A_2-x_2A_1}{AD^2} \rangle =
\frac{(x_1-x_2)L_0}{xx_1x_2(1-x)}\, , \qquad
\langle \frac{1}{D^2} \rangle =\frac{L_0}{xx_1x_2}\, , \\ \nonumber
&& \langle \frac{1}{AD} \rangle = \frac{-L_0}{x_1x_2(1-x)}
\bigl[\frac{1}{2}L_0 + \ln\frac{x_1x_2}{x} \bigr], \qquad
\langle \frac{1}{C^2D} \rangle = \frac{-L_0}{x_1(1-x_2)^3}\, , \\ \nonumber
&& \langle \frac{1}{AC} \rangle = \frac{-L_0}{x_1x_2^2}
\bigl[L_0 + 2\ln\frac{x_1x_2}{x} + 2\ln\frac{xx_2}{(1-x)(1-x_2)}
\bigr], \qquad
\langle \frac{1}{A^2D} \rangle = \frac{-L_0}{(1-x)^3}, \\  \nonumber
&& \langle \frac{A}{C^2D^2} \rangle = \frac{x_2L_0}{x_1(1-x_2)^4}, \qquad
\langle \frac{C}{A^2D^2} \rangle = \frac{-x_2L_0}{(1-x)^4}\, , \\ \nonumber
&& \langle \frac{A}{CD^2} \rangle = \frac{-L_0}{x_1(1-x_2)^2}
\bigl[ \frac{1}{2}L_0 + \ln\frac{x_1x_2}{x} \bigr]
+ L_0\frac{x_2x-x_1}{xx_1x_2(1-x_2)^2}\, , \\  \nonumber
&& \langle \frac{C}{AD^2} \rangle = \frac{-L_0}{x_1(1-x)^2}
\bigl[ \frac{1}{2}L_0 + \ln\frac{x_1x_2}{x} \bigr]
- L_0\left( \frac{x_1-x_2}{x_1x_2(1-x)^2} + \frac{1}{xx_2(1-x)} \right)\, .
\end{eqnarray}

\section*{Appendix B}

\setcounter{equation}{0}
\renewcommand{\theequation}{B.\arabic{equation}}

Here we derive eq.~(\ref{eq:28}), starting from
eq.~(\ref{eq:38}), by integration over regions (\ref{eq:26}) and
(\ref{eq:27}), taking into account the aperture of the detectors.
Let us note first that
\begin{eqnarray} \label{eq:b1}
&& \frac{\dd (q_2^{\bot})^2}{(q_2^{\bot})^2}\cdot
\frac{\dd (q_1^{\bot})^2}{(q_1^{\bot}+q_2^{\bot})^2}\cdot
\frac{\dd \phi}{2\pi(q_1^{\bot}+xq_2^{\bot})^2}
=\frac{1}{Q_1^2}\cdot\frac{\dd z_2}{z_2}\cdot
\frac{\dd z_1 \dd \phi}{2\pi(1-x)(z_2x-z_1)}
 \nonumber \\ && \qquad \cdot
\biggl[-\frac{1}{z_1+z_2+2\sqrt{z_1z_2}\cos\phi}
+\frac{1}{(z_1/x)+xz_2+2\sqrt{z_1z_2}\cos\phi} \biggr].
\end{eqnarray}

Integrating the right-hand side of eq.~(\ref{eq:b1}) we have to keep in mind
that
the first term in the brackets is sensitive to region (\ref{eq:26})
and the second to region (\ref{eq:27}). The aperture of the symmetrical
circular detector is shown in fig.~3.

%-----------------fig.3-----------------------------------
%\begin{figure}
\vspace{.7cm}
\unitlength=1.0mm
\special{em:linewidth 0.6pt}
\linethickness{0.6pt}
\begin{picture}(150.00,85.00)
\put(10.00,10.00){\vector(0,1){70.00}}
\put(13.00,81.00){\makebox(0,0)[cc]{$z_1$}}
\put(10.00,10.00){\vector(1,0){135.00}}
\put(146.00,13.00){\makebox(0,0)[cc]{$z_2$}}
\put(30.00,20.00){\line(0,1){50.00}}
\put(30.00,20.00){\line(1,0){100.00}}
\put(130.00,20.00){\line(0,1){50.00}}
\put(30.00,70.00){\line(1,0){100.00}}
\put(10.00,10.00){\line(2,1){130.00}}
\put(140.00,78.00){\makebox(0,0)[cc]{$z_1=xz_2$}}
\put(10.00,10.00){\line(1,1){70.00}}
\put(85.00,76.00){\makebox(0,0)[cc]{$z_1=z_2$}}
\put(30.00,28.00){\line(1,1){42.00}}
\put(30.00,32.00){\line(1,1){38.00}}
\put(32.00,20.00){\line(2,1){98.00}}
\put(30.00,21.00){\line(2,1){98.00}}
\put(27.00,30.00){\makebox(0,0)[cc]{$2\delta\{$}}
\put(90.00,40.00){\vector(0,1){09.00}}
\put(90.00,60.00){\vector(0,-1){09.00}}
\put(95.00,42.00){\makebox(0,0)[cc]{$2x\delta$}}
\put(10.00,20.00){\line(1,0){3.00}}
\put(15.00,20.00){\line(1,0){3.00}}
\put(20.00,20.00){\line(1,0){3.00}}
\put(25.00,20.00){\line(1,0){3.00}}
\put(07.00,20.00){\makebox(0,0)[cc]{$x^2$}}
\put(10.00,70.00){\line(1,0){3.00}}
\put(15.00,70.00){\line(1,0){3.00}}
\put(20.00,70.00){\line(1,0){3.00}}
\put(25.00,70.00){\line(1,0){3.00}}
\put(05.00,70.00){\makebox(0,0)[cc]{$x^2\rho^2$}}
\put(30.00,10.00){\line(0,1){3.00}}
\put(30.00,15.00){\line(0,1){3.00}}
\put(30.00,07.00){\makebox(0,0)[cc]{$1$}}
\put(130.00,10.00){\line(0,1){3.00}}
\put(130.00,15.00){\line(0,1){3.00}}
\put(130.00,07.00){\makebox(0,0)[cc]{$\rho^2$}}
\end{picture}
%\caption{ The aperture of the symmetrical circular detector
%for the integration over $z_1$ and $z_2$ in the case where
%only the initial electron loses energy for the pair creation;
%$\delta=2\sqrt z_2\lambda(1-x)$.}
%\end{figure}

\nopagebreak
\vspace{.1cm}
\noindent
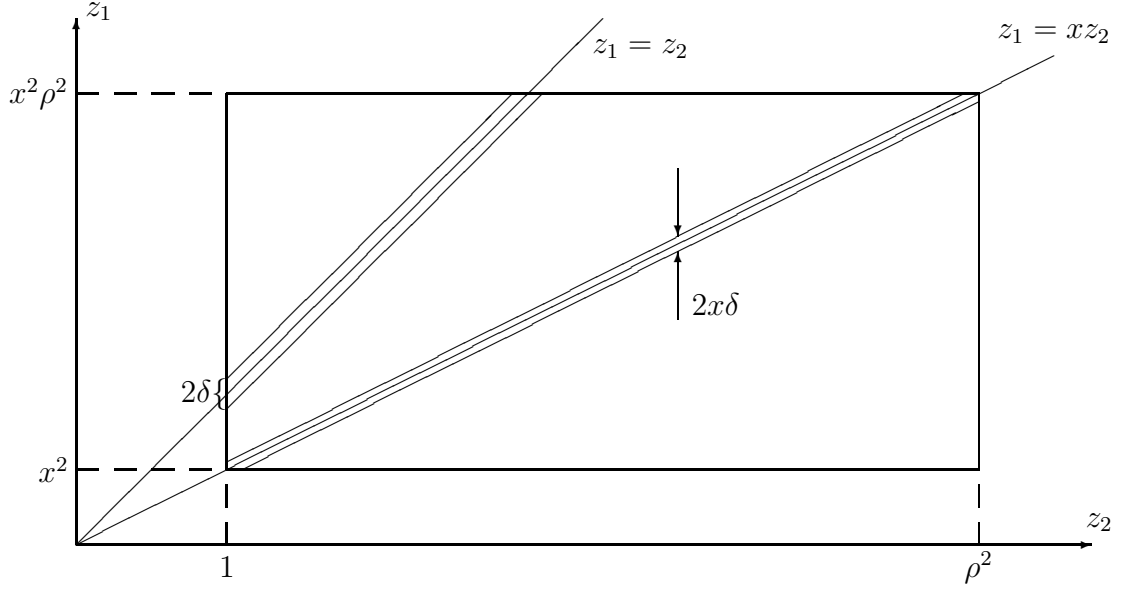
Fig.~3. {\small The aperture of the symmetrical circular detector
for the integration over $z_1$ and $z_2$ in the case when
only the initial electron loses energy for the pair creation;
$\delta=2\sqrt{z_2}\lambda(1-x)$.}
\vspace{.3cm}

We present in detail only the integration of the first term in the brackets
in the right-hand side of eq.~(\ref{eq:b1}).
If $|\sqrt{z_1}-\sqrt{z_2}|<\lambda(1-x)$ then the angular integration gives
\begin{eqnarray} \label{eq:b2}
\frac{1}{2\pi}\int\!\!\! \frac{\dd \phi}{z_1+z_2+2\sqrt{z_1z_2}\cos\phi}
&=& \frac{2}{\pi} \int\limits_{0}^{\phi_{\mbox{\small max}}}\!\!\!\frac{\dd
\phi}
{z_1+z_2+2\sqrt{z_1z_2}\cos\phi} \\ \nonumber
&=& \frac{2}{\pi\sqrt{a^2-b^2}}\,\mbox{arctan}\left.
\bigl(\frac{a-b}{\sqrt{a^2-b^2}}\tan\frac{\phi}{2}\bigr)
\right|_0^{\phi_{\mbox{\small max}}} \, ,
\end{eqnarray}
where
\begin{eqnarray*}
\phi_{\mbox{\small max}}&=&\mbox{arccos}\bigl(-1
+\frac{\lambda^2(1-x)^2-(\sqrt{z_1}-\sqrt{z_2})^2}{2\sqrt{z_1z_2}}\bigr),
\\ \nonumber
a&=&z_1+z_2, \qquad b=2\sqrt{z_1z_2}.
\end{eqnarray*}
Because of the smallness of the values $\lambda^2(1-x)^2$ and
$|\sqrt{z_1}-\sqrt{z_2}|$ with respect to $z_1$ and $z_2$,
we can rewrite the last term in eq.~(\ref{eq:b2}) in the following form:
\begin{eqnarray}
J=\frac{1}{\pi}\cdot \frac{1}{\sqrt{z_2}\,|\sqrt{z_1}-\sqrt{z_2}|}
\,\mbox{arctan}\frac{|\sqrt{z_1}-\sqrt{z_2}|}
{\sqrt{\lambda^2(1-x)^2-(\sqrt{z_1}-\sqrt{z_2})^2}}\, .
\end{eqnarray}
Let $z_2>z_1$, then in the region under consideration we have
$\sqrt{z_1}>\sqrt{z_2}-\lambda(1-x)$, and we can carry out
the subsequent integration over $z_1$ in eq.~(\ref{eq:b1})
by taking $z_1=z_2$ in the factor $(xz_2-z_1)^{-1}$ and
introducing the new variable $t=\lambda(1-x)(\sqrt{z_2}-\sqrt{z_1})$
in $J$. Thus we obtain:
\begin{eqnarray}
J=2\,\frac{2}{\pi}\int\limits_{0}^{1}\!\!\!\;\frac{\dd t}{t}\,\mbox{arctan}
\frac{t}{\sqrt{1-t^2}}=2\ln2,
\end{eqnarray}
where the additional factor $2$ is due to the contribution
when $z_1>z_2$. From fig.~3 we see that the region
$|\sqrt{z_1}-\sqrt{z_2}|<\lambda(1-x)$ contributes only
if $z_2<x^2\rho^2$. That is why we have to write
the contribution corresponding to $|\sqrt{z_1}-\sqrt{z_2}|<\lambda(1-x)$
in eq.~(\ref{eq:b1}) as:
\begin{eqnarray}
\frac{1}{Q_1^2}\int\limits_{1}^{\rho^2}\!\!\!\;\frac{\dd z_2}{z_2^2}
\frac{1}{(1-x)^2}\, 2\Theta\ln2, \qquad \Theta=\Theta(x^2\rho^2-z).
\end{eqnarray}

If now $|\sqrt{z_1}-\sqrt{z_2}|>\lambda(1-x)$ the angular integration
is trivial and the subsequent integration over $z_1$ and $z_2$ is reduced
to the integration of the function $\{(z_1-xz_2)|z_1-z_2|\}^{-1}$
over the rectangle $1<z_2<\rho^2$, $\ x^2<z_1<x^2\rho^2$ without the narrow
strip of width $2\delta$ $(\delta=2\sqrt{z_2}\lambda(1-x))$.
The result reads:
\begin{eqnarray} \label{eq:b6}
&& \frac{1}{Q_1^2}\int\limits_{1}^{\rho^2}\frac{\dd z_2}{z_2^2}
\frac{1}{(1-x)^2}\, \biggl\{ \ln\left|\frac{(z_1-x^2)(x\rho^2-z_2)}
{(x-z_2)(z_2-x^2\rho^2)}\right| + \Theta\,\biggl(
-2\ln2 \\ \nonumber && \qquad
+ \ln \frac{(x^2\rho^2-z_2)^2}{(x\rho^2-z_2)^2x^2} \biggr) \biggr\}\, .
\end{eqnarray}
It easy to see that the full contribution of the first term
in brackets in eq.~(\ref{eq:b1}) is reduced to eq.~(\ref{eq:b6})
without $-2\Theta\ln2$ in the latter.

The integration of the second term in brackets in eq.~(\ref{eq:b1})
can be done in full analogy. The result could be written as:
\begin{eqnarray}
\frac{1}{Q_1^2}\int\limits_{1}^{\rho^2}\!\!\!\;\frac{\dd z_2}{z_2^2}
\frac{1}{(1-x)^2}\, \biggl\{ \ln \frac{z_2}{\lambda^2}
+ \ln\left|\frac{(\rho^2-z_2)(z_2-1)}{(x-z_2)(z_2-x\rho^2)}
\right| \biggr\}\, ,
\end{eqnarray}
and formula (\ref{eq:28}) becomes obvious.

\end{document}